\documentclass[prd, twocolumn, nofootinbib, floatfix, notitlepage,longbibliography]{revtex4-2}

\usepackage{amsmath}
\usepackage{graphicx}
\usepackage{dcolumn}
\usepackage{bm}
\usepackage{epsfig}
\usepackage{amssymb,latexsym,mathrsfs}
\usepackage{graphicx}
\usepackage{color}
\usepackage{xcolor}
\usepackage{hyperref}

\hypersetup{
    colorlinks=true,
    linkcolor=red,
    citecolor=blue,
}

\newcommand{\be}{\begin{equation}}
\newcommand{\ee}{\end{equation}}
\newcommand{\bs}{\begin{split}} 
\newcommand{\bea}{\begin{eqnarray}}
\newcommand{\eea}{\end{eqnarray}}

\newcommand{\om}{\Omega_m}

\newcommand{\oma}{\Omega_m(a)}

\newcommand{\ode}{\Omega_{\rm de}}
\newcommand{\lcdm}{$\Lambda$CDM} 
\newcommand{\gm}{G_m}

\newcommand{\keff}{k_{\rm eff}} 
\newcommand{\kmax}{k_{\rm max}}

\newcommand{\fs}{f\sigma_8} 
\newcommand{\fsh}{\widehat{f\sigma_8}} 
\newcommand{\mds}{M_{\rm dS}} 
\newcommand{\dg}{\delta G} 
\newcommand{\sigo}{\sigma_{8,0}}
\providecommand\mathplus{+}

\begin{document}

\title{Constraining Scale Dependent Growth with Redshift Surveys} 

\author{Mikhail Denissenya${}^{1}$, Eric V. Linder${}^{1,2}$} 
\affiliation{
${}^1$Energetic Cosmos Laboratory, Nazarbayev University, Nur-Sultan, 
Qazaqstan 010000\\ 
${}^2$Berkeley Center for Cosmological Physics \& Berkeley Lab, 
University of California, Berkeley, CA 94720, USA
} 

\begin{abstract}
Ongoing and future redshift surveys have the capability to 
measure the growth rate of large scale structure at the 
percent level over a broad range of redshifts, tightly  
constraining cosmological parameters. Beyond general 
relativity, however, the growth rate in the linear density 
perturbation regime can be not only redshift dependent but 
scale dependent, revealing important clues to modified gravity. 
We demonstrate that a fully model independent approach of binning  
the gravitational strength $G_{\rm eff}(k,z)$ matches  
scalar-tensor results for the growth rate $f\sigma_8(k,z)$ to 
$0.02\%$--$0.27\%$ rms accuracy. For data of the quality of 
the Dark Energy Spectroscopic Instrument (DESI) we find the 
bin values can be constrained to 1.4\%--28\%. We also 
explore the 
general scalar-tensor form, constraining the amplitude and 
past and future scalaron mass/shape parameters. 
Perhaps most interesting is the strong complementarity 
of low redshift peculiar velocity data with DESI-like 
redshift space distortion measurements, enabling 
improvements up to a factor 6--7 on 2D joint confidence 
contour areas. 
Finally, we quantify some issues with gravity 
parametrizations that do not include all the key physics. 
\end{abstract} 

\date{\today} 

\maketitle


\section{Introduction} \label{sec:intro} 

Cosmic growth of large scale structure is a competition 
between increased clustering amplitude driven by gravity 
and damped evolution due to expansion, exacerbated by 
cosmic acceleration. Thus the growth, and more incisively 
the growth rate, can reveal important characteristics of 
gravitation and dark energy. The new generation of spectroscopic 
galaxy surveys measure redshift space distortions across 
wide areas of sky and broad redshift ranges, enabling 
precision estimates of the growth rate $\fs$. 

Within general relativity these measurements translate 
directly to cosmic expansion history constraints, but 
the extra freedom within modified gravity means that 
redshift space distortions can serve as a critical probe 
of gravity (see, e.g., 
\cite{0709.1113,0802.1944,0803.2236,0807.0810} for early work). 
Moreover, with modified gravity there generally enters 
scale dependence in the growth rate, even in the linear 
density perturbation regime, a further signal of deviation 
from Einstein gravity. 
We focus here on constraining 
scale dependent growth, in as model independent a manner 
as practical, with the quality of data likely to be 
delivered by the current spectroscopic surveys such as 
from the Dark Energy Spectroscopic Instrument (DESI, 
\cite{1611.00036,2205.10939}). 
(While scale dependence can enter due to massive neutrinos, this has a specific scale and redshift dependence that can be fit, while scale dependence from, e.g., a warm dark matter component generally enters beyond the linear scales for viable models. Therefore we only explore scale dependence from modified gravity here.) 

In Section~\ref{sec:scale} we investigate the evolution 
of the scale- and redshift-dependent cosmic growth rate 
$\fs(k,a)$ and explore the impact of the physics motivated 
scalar-tensor gravity form for the modified gravity strength, 
keeping it as model independent as reasonable. 
Section~\ref{sec:binopt} then demonstrates that a wholly 
model independent form, using $2\times3$ 
bins in scale and redshift, 
can provide an excellent approximation to the full form. 
We project constraints in Section~\ref{sec:infomatrix} 
for what DESI-quality data could enable to test gravity 
through scale dependent growth in these two approaches. 
Leverage from peculiar velocity low redshift observations 
is examined as well. Section~\ref{sec:bin22} studies 
a coarser, but easier to be alerted to distinctions 
from general relativity, $2\times2$ bin approach. 
In Section~\ref{sec:concl} we 
summarize and conclude.

\section{Scale Dependent Growth} \label{sec:scale} 

Galaxy redshift surveys use redshift space distortions to 
determine the growth rate factor 
\be 
\fs(a)=fD\,\frac{\sigo}{D_0}=\frac{dD}{d\ln a}\,\frac{\sigo}{D_0}\ , 
\ee 
where $D$ is the linear perturbation growth factor, 
$a$ is the expansion factor, and $\sigma_8$ the 
mass fluctuation amplitude; subscript 0 denotes 
the value today. Unlike for the logarithmic growth 
rate $f=d\ln D/d\ln a$, the evolution equation for 
$D$ is linear: 
\be 
D''+D'\left[2+\frac{1}{2}\left(\ln H^2\right)'\right]-\frac{3}{2}D\gm\oma=0\ , 
\ee 
where prime denotes a derivative with respect to $\ln a$. 

It is convenient to break this second order equation 
into two coupled first order equations, 
\bea 
D'&=&(fD)\\ 
(fD)'&=&-(fD)\left[2+\frac{1}{2}\left(\ln H^2\right)'\right]+\frac{3}{2}D\gm\oma\ . \label{eq:fdevo} 
\eea  
Here $H(a)$ is the Hubble parameter $\dot a/a$, 
and $\oma$ is the matter density in units of the 
critical density. The factor $\gm$ is the 
gravitational coupling strength of matter, in 
units of Newton's constant. In general relativity 
it is simply one. 

However, in modified gravity $\gm$ can differ 
from one, and be both expansion factor (redshift) 
dependent and scale (wavemode) dependent. 
Since we work in the linear regime, linear 
perturbations at one wavenumber $k$ are independent 
of those at another. In general then we have 
$\gm(k,a)$, determining $\fs(k,a)$.

\subsection{Modified Gravity} \label{sec:modgr} 

One could adopt a particular theory of modified 
gravity, with a certain functional form, and 
certain parameters within that form -- e.g.\ 
$f(R)$ gravity of the Hu-Sawicki form with 
parameter $n=4$ -- but the constraints derived from 
the data will then apply only to that specific case. 
We take a more model independent approach, using 
that in scalar-tensor theories the gravitational 
coupling to matter can be fairly generically 
written as \cite{0511218,0611321,0709.0296,0801.2431,0809.3791,1108.4242,1302.1193,1409.8284} 
\be 
\gm=\frac{1+c\,\left(k/[aM(a)]\right)^2}{1+\left(k/[aM(a)]\right)^2}\ . \label{eq:pade} 
\ee 
This comes directly from the Einstein field equations 
in the quasistatic regime, where precision 
observations can be made, and can be seen as well in both the effective field theory of dark energy approach \cite{1606.05339} and property function approach, e.g.\ \cite{2210.01094}. 
We take the scalaron mass to be given by 
\be 
M(a)=M_a a^{-3}+\mds\ , \label{eq:mofa}
\ee 
which fits many models reasonably 
\cite{1103.0282} (also see \cite{1101.0295}), 
and guarantees a high redshift universe looking 
like that in general relativity. 
Since in general relativity $\gm=1$, 
we define $\dg=\gm-1$. 
We refer to $c$ as the amplitude parameter 
and $M_a$ and $\mds$ as the shape parameters, 
since they govern the evolution of scale dependence. 
Note that at high redshift the scalaron mass 
is large, $M\approx M_a a^{-3}$, today 
$M(a=1)=M_a+\mds$, and in the future the 
scalaron mass freezes as the universe approaches 
a de Sitter state, $M(a)\to\mds$. Since today 
the expansion is not too far from the de Sitter 
state, i.e.\ $\ode\approx0.7$, then we expect 
$M(a=1)=M_a+\mds$ to be not far from $\mds$. Thus $M_a\lesssim\mds$ is a reasonable value. 

Consider a fiducial model of $c=4/3$, $M_a=0.05$, 
$\mds=0.05$. 
(All wavenumbers and masses are in $h$/Mpc units.) 
The value $c=4/3$ matches $f(R)$ gravity \cite{0809.3791}, 
and the value $M(a=1)=M_a+\mds=0.1$ gives a value 
for the Compton wavelength parameter \cite{0705.1158} 
$B_0=2[H_0/M(a=1)]^2\approx 2.2\times 10^{-5}$. 
The exact translation of $B_0$ to the $f(R)$ parameter 
$f_{R0}$ is model dependent, but generally 
$f_{R0}\approx -B_0/(1-4)$. Thus the fiducial model 
has $f_{R0}\approx -10^{-5}$, toward the high end of 
what is allowed by current data, designed to stress 
test the ability of our binned approach to accurately 
fit the growth data. Finally note that the power 
$a^{-3}$ in Eq.~\eqref{eq:mofa} also stress tests 
our approach in that this is the shallowest power 
allowed: one requires $M(a\ll1)\sim a^{\le-3}$ \cite{0809.3791,1008.2693} 
to obtain the standard matter dominated 
Friedmann expansion equation behavior. Using the 
limit $M(a\ll1)\sim a^{-3}$ keeps modifications stronger 
at earlier times and so makes fitting growth accurately 
with a small number of $\dg$ bins more difficult.

\begin{figure*}	
\includegraphics[width=0.48\textwidth]{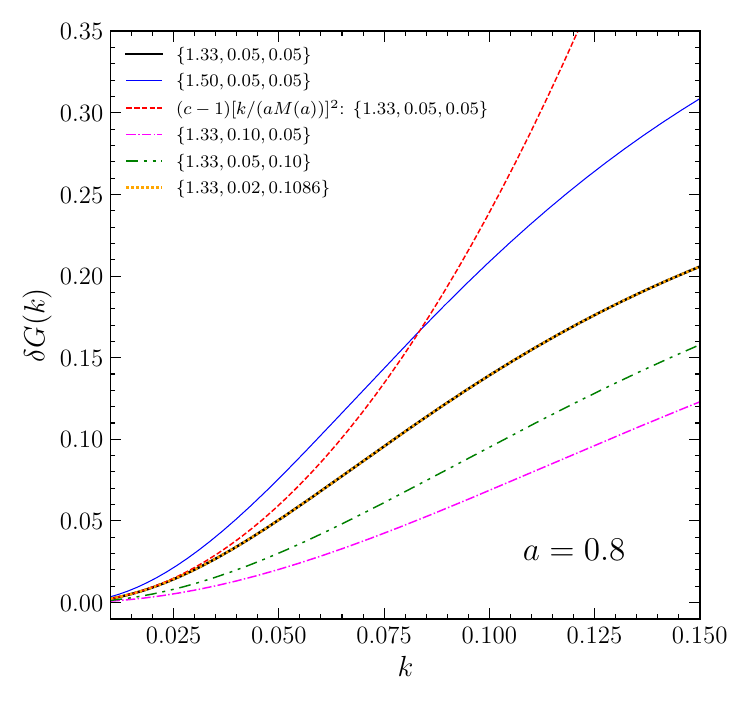}
\includegraphics[width=0.48\textwidth]{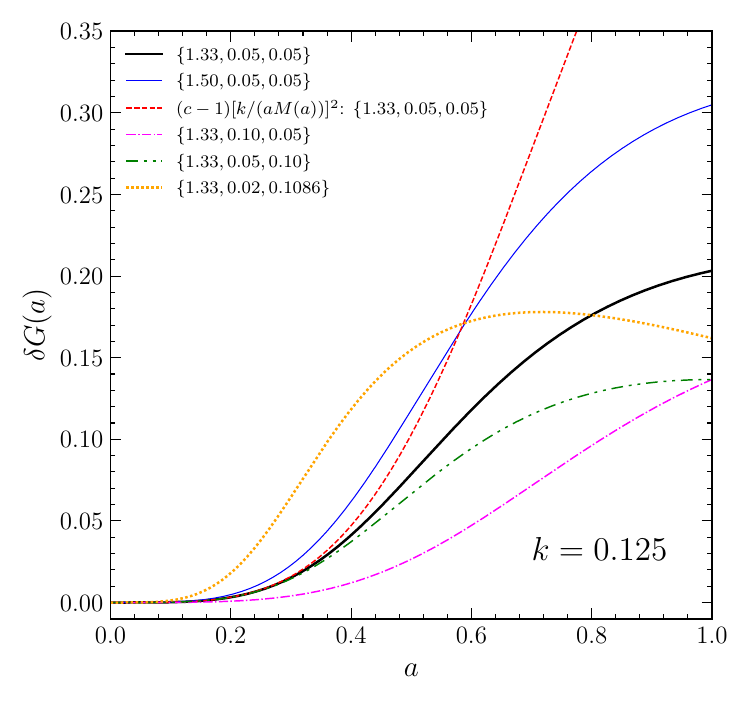}   
\caption{
Modified gravitational strength deviations from GR are 
plotted as $\dg$ vs $k$, at $a=0.8$ (left panel), and 
$\dg$ vs $a$ at $k=0.125$ (right panel). The fiducial 
case of $c=4/3$, $M_a=\mds=0.05$ (bold black curve) and 
several variations as labeled by $(c,M_a,\mds)$ are shown, 
as well as an unmotivated power law rather than Pad\'e polynomial case (dashed red). 
}
\label{fig:dG}
\end{figure*}

Figure~\ref{fig:dG} shows $\dg(k)$ for $a=0.8$, i.e.\ 
in the recent universe where the deviation from general relativity 
(GR) is larger, and $\dg(a)$ for $k=0.125$, i.e.\ at a higher 
wavenumber where the deviation is larger. We illustrate 
the effects of changing the parameters one by one. 
Increasing the amplitude parameter $c$ of course scales 
$\dg$ at all $k$ and $a$ (compare thin blue vs bold 
black curves). For the shape parameters entering the 
scalaron mass, recall that large mass increases the 
similarity to GR. Increasing the shape parameter $M_a$ 
greatly suppresses modifications until recently, 
for example we see that $\dg$ is three times smaller 
at $a=0.5$ (for $k=0.125$) for $M_a=0.1$ rather than 
$M_a=0.05$ (compare dot dash purple vs bold black curves). 
The shape parameter $\mds$ changes the late time behavior; 
it is not until $a\gtrsim0.5$ that increasing $\mds$ from 
0.05 to 0.1 has an appreciable effect (dot dot dash green 
vs bold black). 

There are two other points of interest to note. First, using 
a form like $\gm=1+ba^s$ 
is quite problematic. See the detailed analysis in \cite{1612.00812}, as well as 
\cite{1902.10503} and the model independent approach of \cite{2207.09896}. It gives 
very different behavior than the full Pad{\'e} ratio of 
Eq.~\eqref{eq:pade}; this is particularly evident in the dashed red curve in 
Fig.~\ref{fig:dG}. 
The basic issue is that it overweights late times 
(and high $k$) so that a constraint from late time data 
could be interpreted as an unrealistically tight limit on 
modified gravity. Recall that the Pad\'e form originates 
from the Einstein field equations 
\cite{0511218,0611321,0709.0296,0801.2431,1108.4242,1302.1193,1409.8284}; basically $\gm$ 
comes from $F_1$(metric potential) = $F_2$(matter fields), 
where $F_i$ are factors including space and time derivative 
terms, and matter fields include the scalar field 
perturbations; both sides involve $k^2$ corrections beyond leading order. Since 
$\gm$ arises from a ratio of the factors, using only a 
single polynomial rather than a Pad\'e form is throwing 
away half the physics. This holds for any scalar-tensor 
theory. We can see from Fig.~\ref{fig:dG} that the 
discrepancy is already significant for $a\gtrsim0.5$ or 
$k\gtrsim0.06$. 

Secondly, as mentioned the scalaron mass 
freezes approaching a dark energy dominated de Sitter 
state, $\gm\not\to c$ but rather gravity is eventually 
restored to GR, $\gm\to1$ since $aM(a)\to a\mds$ and 
the factors multiplying the $k^2$ terms go to zero. The 
``phase space'' evolution of $\gm$ for various modified 
gravity theories is discussed and illustrated in 
\cite{1103.0282}. Whether $\gm$ reaches its fully 
modified value of $c$ before turning around and 
approaching back to 1 depends on the values of 
$M_a$ and $\mds$. We illustrate this in Fig.~\ref{fig:dG} 
with the dotted orange curve, where $M_a$ and $\mds$ 
are chosen to match $M(a=0.8)$ from the fiducial 
model (specifically, we use $M_a$=0.02, $\mds=0.1086$). 
Indeed, the left panel shows perfect overlap 
between those two cases, while the right panel shows 
that they coincide only at (the chosen) $a=0.8$. The 
turnaround is clearly seen: the dotted orange curve 
never exceeds $\dg\approx0.18$, well short of $c-1=0.33$. 
All of the curves (except the non-Pad\'e one) will 
eventually turn around; it simply occurs later for 
smaller $\mds$.

\subsection{Growth Rate} \label{sec:growth} 

\begin{figure*}	
\includegraphics[width=\columnwidth]{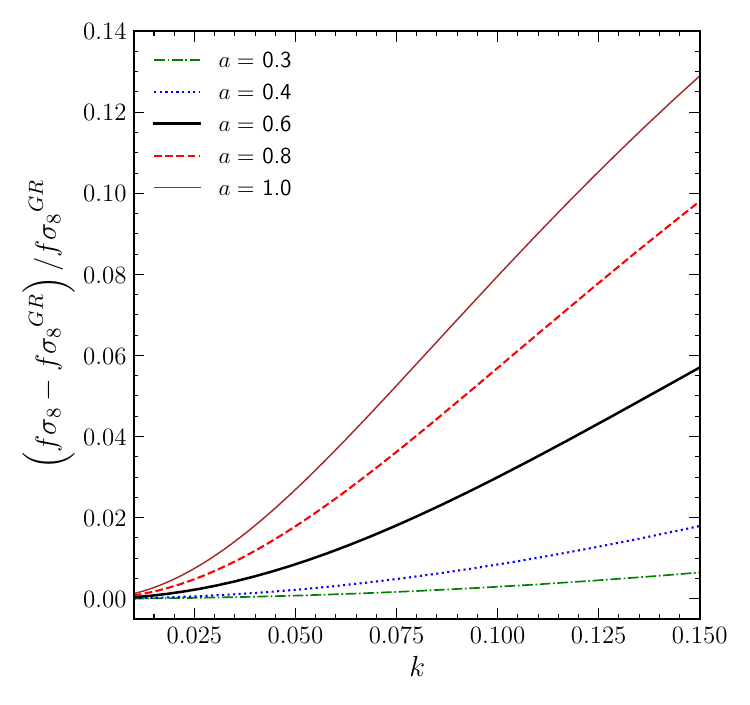}
\includegraphics[width=\columnwidth]{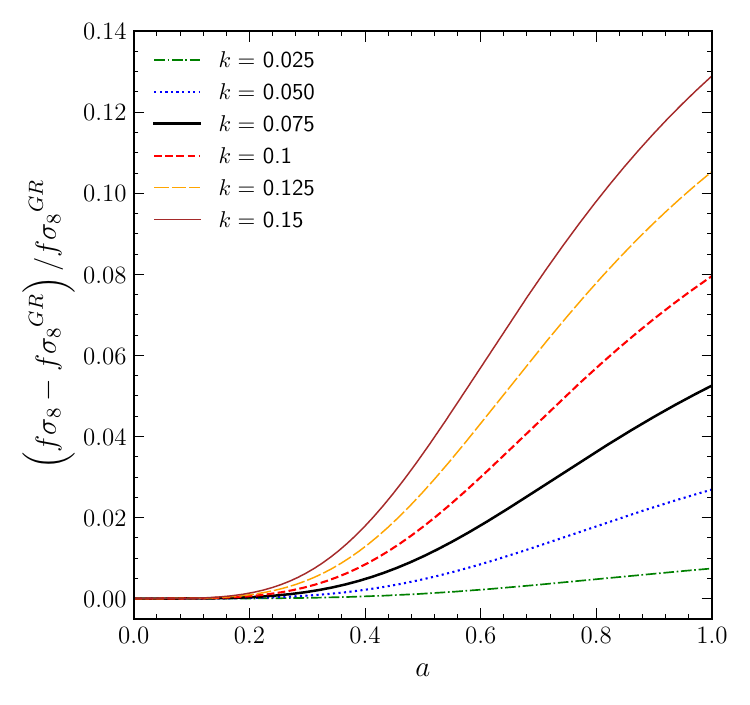}
\caption{The fractional difference of 
$\fs$ relative to general relativity is plotted vs $k$, for various scale factors $a$ [left panel], and vs $a$ for various  wavemodes $k$ [right panel]. 
}
\label{fig:dfsgr}
\end{figure*}

Given $\gm$, Eq.~\eqref{eq:pade}, one can then 
solve Eq.~\eqref{eq:fdevo} and obtain the scale dependent 
growth rate $\fs(k,a)$ in the linear regime. 
Figure~\ref{fig:dfsgr} shows $(\fs-\fs^{GR})/\fs^{GR}$ 
vs $a$ for various $k$, and vs $k$ for various $a$, 
for the fiducial modified 
gravity model, in a \lcdm\ background with fractional 
matter density today $\om=0.3$. 
(Recall that 
the maximum equation of state deviation from 
$-1$ for many $f(R)$ theories is 
$|1+w|_{\rm max}\approx B_0/2\ll 1$ \cite{0905.2962}, 
so \lcdm\ is an excellent approximation.) We normalize 
the amplitude to the same primordial density 
amplitude as GR (as measured by the cosmic microwave 
background), rather than to the present mass 
fluctuation amplitude $\sigo$ (which would cause 
deviations at high redshift).

The growth rate $\fs$ has less than 1\% deviation from 
GR for the fiducial model earlier than about one efold before 
the present, $a\lesssim0.37$, for $k<0.15$. Lower wavenumbers 
deviate later and with a smaller amplitude. Between $a=0.4$ 
($z=1.5$) and $a=0.8$ ($z=0.25$), the deviation grows  
by a factor of $\sim6$. Thus redshift space distortion 
data over a broad range of redshifts is highly useful. 
The deviation from GR in the growth rate increases with $k$, 
reaching 2--10\% at $k=0.15$ for $a=0.4$--0.8, the range 
of near term precision redshift space distortion data. 
Thus we see that percent level measurements of $\fs$ 
will be required, somewhat eased by adding up measurements 
at many redshifts. It is important to remember that 
$\fs(k,a)$ is affected by all values $\gm(k,a'\le a)$; 
i.e.\ while modes evolve independently within the linear 
regime, modifications in the gravitational strength at 
earlier times affect the growth rate at later times 
(effectively changing the ``momentum'' of the growth, 
see \cite{1703.00917}).

\section{Model Independent Fitting} \label{sec:binopt} 

Within scalar-tensor theories, we then have a three 
parameter fit $(c,M_a,\mds)$ for gravity that can be 
used for testing GR with growth rate data. This is a 
valid and useful approach, but while it is the most 
physical approach without adopting a specific gravity 
model it does have some drawbacks. One must solve the 
growth equation for each point in the parameter space, 
and there is some degeneracy. High values of $M(a)$, 
coming from either large $M_a$ or large $\mds$, are 
indistinguishable from GR (for a Monte Carlo fitting 
approach one might want to adopt 
priors on $1/M_a$, $1/\mds$ instead). Covariance 
between the parameters can only partially be ameliorated 
with data over a range of $k$ and $a$. Finally, while 
the Pad\'e form is model independent within canonical  
scalar-tensor theories, we might want to test the 
framework itself. 

Therefore we will also investigate an even more 
model independent approach, that of independent 
bin values of $\gm$ in $k$ and $a$. 
(One can also use node values for $\gm$ and interpolate between them, e.g.\ with linear or spline functions. This smoothness increases the ease of attaining a good fit to the exact form (see Sec.~\ref{sec:bin22}) but also increases the covariance between bins and hence the interpretation of $\gm(z)$. In the spirit of ``stress testing'' the ability of bins to accurately give $\fs$ we use $\gm$ constant in bins.) 
We want to keep 
the number of bins small, but sufficient to both 
see the physics qualitatively and reproduce the 
growth rate with quantitative accuracy. Note that the 
bins are in $\gm$, giving the full, continuous 
function $\fs(k,a)$. 
(For some studies using gravity binned 
in scale and redshift, see 
\cite{1003.0001,1008.0397,1103.1195,1212.0009,1312.1022,1707.06627,1707.08964,1908.00290,2005.14351,2007.12607,2010.12519,2101.12261}.)  

Since we restrict ourselves to the quasilinear regime, 
we choose two bins in $k$: low $k=[0.01,0.1]$ and 
high $k=[0.1,0.15]$. Subdividing further would lead to 
increased covariance between parameters and weaker 
constraints. For expansion factor $a$, we recognize 
that models such as $f(R)$ have very rapid variation 
in their time dependence, e.g.\ $\dg\sim a^4$ at 
low $k$, low $a$. This led \cite{mgbinlo} to adopt 
three bins in $a$ for accuracy, and we follow this 
procedure. We take bins, in redshift, of $z=[0,0.5]$, 
[0.5,1], and [1,3]. Again we emphasize that these are 
bins for $\gm$ and not in $\fs$. Although we saw that 
$\fs$ deviates from GR by less than 1\% for $a<0.37$, 
we must allow $\gm$ to deviate earlier since those 
deviations affect all later $\fs$, even up to $z=0$. 
From Fig.~\ref{fig:dG} we see that $\dg$ deviates 
by less than 1\% at $a<0.25$, and set $\gm=1$ at 
$a<0.25$. 
Note that while 
$\gm$ is piecewise constant in bins, as 
we never take derivatives of $\gm$ then jumps at the 
bin boundaries do not cause difficulties for the 
calculation; smoothing the transitions was studied 
in \cite{mgbinlo} and found not to affect substantially 
the results. 

From the $2\times3$ binned values of $\gm$ we 
compute $\fsh(k,a)$ and compare this to $\fs$ 
from the full, Pad\'e functional form of $\gm$. 
Minimizing the variance of the fractional deviation 
$(\fsh-\fs)/\fs$ determines the optimal values 
for the binned $\gm$. Note that while each $k$ bin is 
independent from the others, since the growth 
equation does not couple modes in the linear regime, 
the evolution of $\fsh$ at some $a$ does depend 
on $\gm$ at earlier $a$ as this affects $D(a)$, 
which has been integrated from the high redshift, 
GR initial conditions to $a$. 

In more detail, we recognize that observational 
data will not have $\fs(k,a)$ at every $k$ with 
sufficient signal to noise (S/N) to test gravity, but 
rather one must assess $\fs$ for modes over some 
range. So as not to decrease the S/N of the data 
too much we choose two bins of $k$ in $\fs$, using 
$k=[0.01,0.1]$ and $k=[0.1,0.15]$. Since a deviation 
$\dg$ at some $k$ leads to a deviation in $\fs$ at 
exactly that $k$ and no other, in the linear regime, 
it is convenient to take the $\fs$ bins in $k$ to 
be the same as the $\gm$ bins in $k$. Measurements 
would be quoted at some $k_{\rm eff}$ in each bin, 
which here we will simply take as the bin center, 
i.e.\ $\keff=0.055$, 0.125. (One could take into 
account the mode density, dependence of the power 
spectrum on $k$, spectrograph fiber collision weights 
\cite{1509.06386}, etc.\ to 
come up with a more sophisticated $\keff$ but we 
simply take the bin center.) In redshift, data is 
usually reported in fairly narrow redshift slices, 
approximating a continuous function for $\fs$ in 
terms of $a$, so we do not bin $\fs$ in $a$ (while 
$\gm$ values are defined in bins in $a$). However, 
the precision of redshift space distortion measurements 
degrades at $z<0.2$ where there is relatively little 
volume (but see Sec.~\ref{sec:lowz} where we revisit 
this using peculiar velocity surveys). At high redshift, 
again the precision worsens due to observational 
difficulties. Therefore we carry out the optimization of 
$\gm(k,a)$ bins by minimizing the variance of deviations 
in $\fs$ over the range $z=[0.2,1.7]$. (Recall 
that $\gm$ at $z>1.7$ still affects $\fs$ over this 
whole range.) 

\begin{figure*}	
\includegraphics[width=\columnwidth]{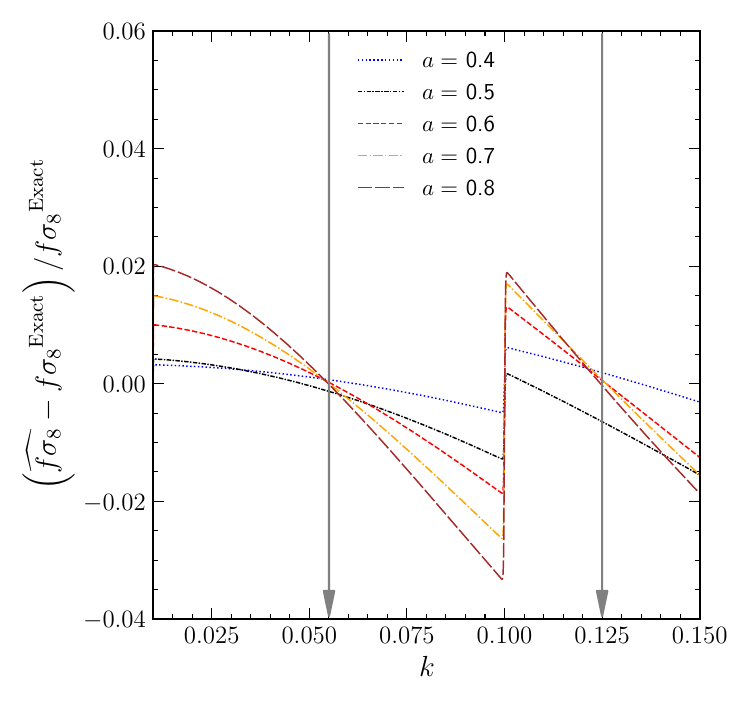}
\includegraphics[width=\columnwidth]{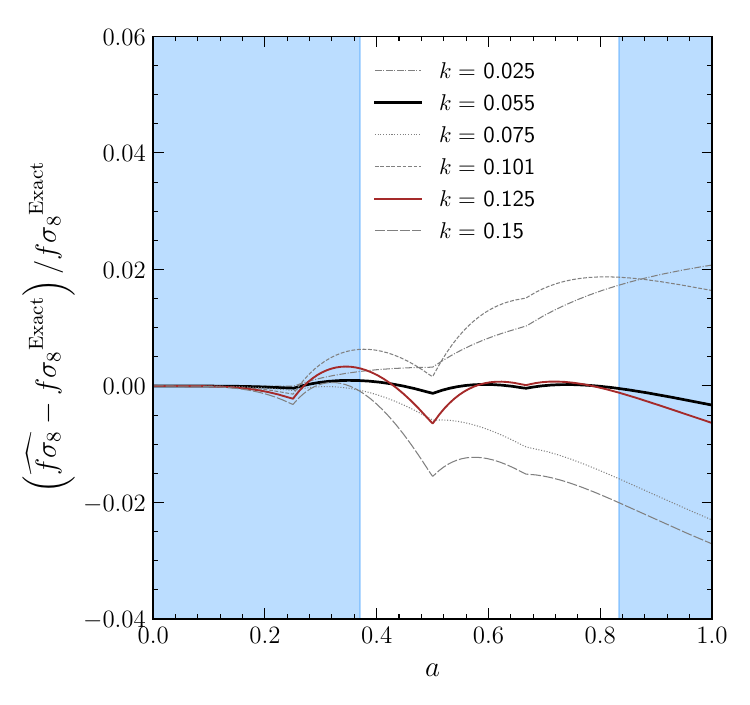} 
\caption{The fractional difference of $\fs$ using the 
optimized binned $\gm^{\rm bin}$ (giving $\fsh$) vs 
using the exact modified gravity $\gm$ 
(giving $\fs^{\rm Exact}$) is plotted vs $k$ for 
various $a$ [left panel] and vs $a$ for various 
$k$ [right panel]. Only the values at $\keff=0.055$, 
0.125 are observationally relevant (since $\fs$ data 
cannot be divided into more than two bins without 
losing too much precision). These are indicated with 
vertical lines in the left panel, and solid curves in 
the right panel. The observationally relevant (i.e.\ good 
data precision) 
range in $a$ is indicated by the unshaded region in 
the right panel. 
	}
\label{fig:dfsgrbe}
\end{figure*}

Figure~\ref{fig:dfsgrbe} shows the fractional residuals 
of the fit $(\fsh-\fs)/\fs)$ vs $k$ for various values 
of $a$, and vs $a$ for various $k$. {\it However\/}, 
what is important are the results at the two values 
$\keff=0.055$ and 0.125 since $\fs(\keff,a)$ is what 
the data will actually provide. These values are marked 
by the vertical lines in the left panel, and the solid 
curves in the right panel. We see that over the redshift 
range of the data, the binned $\gm$ approach succeeds in 
delivering accurate results, with the maximum deviation $\lesssim0.1\%$ in the low 
$k$ bin, and $\lesssim0.1\%$ in the high  
$k$ bin except right around $a=0.5$ (a $\gm$ bin boundary) 
where the maximum deviation is $\approx0.6\%$. Changing from 
piecewise constant $\gm$ to a smoother form would 
reduce this deviation, but 0.6\% is still sufficiently 
good as we will see the observational precision at this 
redshift is expected to be $\approx2\%$ when the redshift 
space distortions are analyzed in two $k$ bins. 
The rms deviations are even smaller, as shown in 
Table~\ref{tab:dfs}.

\begin{table} 
\centering 
\begin{tabular}{l|c|c} 
\hline 
Range & rms & max \\ 
\hline 
\rule{0pt}{1.05\normalbaselineskip}low $k$, low $a$\quad{} & {}\quad\, 0.07\% \quad\, & {}\quad\, 0.13\% \quad\, \\ 
\rule{0pt}{1.05\normalbaselineskip}low $k$, mid $a$\quad & 0.04\% & 0.13\% \\ 
\rule{0pt}{1.05\normalbaselineskip}low $k$, high $a$ & 0.02\% & 0.04\% \\ 
\rule{0pt}{1.05\normalbaselineskip}high $k$, low $a$ & 0.27\% & 0.64\% \\ 
\rule{0pt}{1.05\normalbaselineskip}high $k$, mid $a$ & 0.22\% & 0.64\% \\ 
\rule{0pt}{1.05\normalbaselineskip}high $k$, high $a$ & 0.06\% & 0.12\% \\ 
\hline 
\end{tabular} \\  
\caption{Fractional deviations on the growth rate $\fs$ 
computed using binned $\gm$ vs the exact form. Low/high 
$k$ refers to $\keff=0.055$, 0.125 respectively; low/mid/high 
$a$ refers to the ranges [0.37,0.5], [0.5,0.667], [0.667,0.833]. 
The rms and max columns give the rms and maximum deviations 
across those ranges of $a$.} 
\label{tab:dfs} 
\end{table}

While in Fig.~\ref{fig:dfsgrbe} we show $(\fsh-\fs)/\fs$ 
at other values of $k$ 
than $\keff$, we again emphasize that the values at 
$\keff$ are what is important when comparing with data. 
The $\gm$ values obtained from our variance minimization are: 
$\dg^{\rm bin}_{11}=0.0065$, $\dg^{\rm bin}_{12}=0.031$, $\dg^{\rm bin}_{13}=0.052$, $\dg^{\rm bin}_{21}=0.029$, $\dg^{\rm bin}_{22}=0.127$, $\dg^{\rm bin}_{23}=0.158$, 
where $\dg^{\rm bin}$ denotes $G_m^{\rm bin}-G^{GR}$ and the numerical 
indices give the bins of $\gm$ in $k$ and $a$ respectively, 
from low 
values of $k$ or $a$ to high. For example, $\dg^{\rm bin}_{23}$ 
is the optimized binned deviation in gravity in the 
$\gm$ bin $k=[0.1,0.15]$, $a=[0.667,1]$. As expected there is 
a clear trend of strengthening gravity for increasing $k$ 
or $a$.

\section{Projected Constraints} \label{sec:infomatrix} 

To estimate parameter constraints on the gravitational 
model  we can employ the information matrix formalism, 
where 
the information matrix is 
\bea 
F_{ij}&=&\sum_{k_{\rm eff,n}} \sum_{z_m} 
\frac{\partial\fs(k_{\rm eff,n},z_m)}{\partial p_i} 
\frac{\partial\fs(k_{\rm eff,n},z_m)}{\partial p_j}\\ 
&\,&\qquad\times \frac{1}{[\sigma(k_{\rm eff,n},z_m)\fs(k_{\rm eff,n},z_m)]^2} 
\ , \label{eq:infomat} 
\eea 
and the marginalized parameter constraints are given by 
the elements of $F^{-1}$, with the uncertainty on a 
parameter equal to the square root of its diagonal 
element, i.e.\ $\sigma(p_i)=\sqrt{(F^{-1})_{ii}}$. 

The set of data $\fs(k_{\rm eff,n},z_m)$ are taken 
to follow those projected for the Dark Energy Spectroscopic 
Instrument (DESI), using only the redshift space distortion 
galaxy data. We include the 
redshift range $z_m=0.25, 0.35,\dots 1.65$ 
where precision is strong, with the  
uncorrelated fractional 
precisions $\sigma(k_{\rm eff,n},z_m)$ given by 
Tables 2.3 and 2.5 of \cite{1611.00036}. We evaluate 
$\fs$ at $\keff=0.055$ and 0.125, reflecting that the 
observed data must extend over a range, here respectively 
$k=[0.01,0.1]$ and $k=[0.1,0.15]$, to obtain good 
precision. For the lower  $k$ bin, we use the column 
in the $\kmax=0.1$ column of those tables; for the higher 
$k$ bin we use 
$\sigma_{[0.1,0.15]}=1.5(\sigma_{\kmax=0.2}^{-2}-\sigma_{\kmax=0.1}^{-2})^{-1/2}$, 
where the 1.5 factor is a reasonable approximation 
(generally it ranges from $\sim1.3-1.7$ depending on redshift and treatment of quasilinearity) 
for the uncertainty for the range $k=[0.1,0.15]$ vs [0.1,0.2]. 
The background cosmology is taken as \lcdm\ with 
$\om$ a parameter to be marginalized over.

\subsection{Scalar-Tensor  Parameter Constraints} \label{sec:pade} 

We first consider the general scalar-tensor form 
Eq.~\eqref{eq:pade}, so our fit parameters are 
$(c,M_a,\mds,\om)$, with fiducial values 
$(4/3,0.05,0.05,0.3)$. Our standard analysis includes 
a prior $\sigma(\om)=0.01$ 
(representing constraints from other experiments apart from on the growth rate) 
but we also examine 
variations of this. 

Results show that the amplitude parameter $c$ 
cannot be constrained, with $\sigma(c)\approx c$ 
even when fixing $\om$. This is due to strong covariance 
with the shape parameters $M_a$ and $\mds$ (correlation 
coefficients $0.994$ and $0.974$ respectively) -- since 
$c$ is both redshift and scale independent, it cannot 
be distinctly separated from the parameters that do 
affect the redshift and scale dependence of $\fs$. 
(For example an increase in $c$ can be compensated by a combination of increases in $M_a$ and $\mds$.) 
Only when those parameters are fixed does the uncertainty 
on the amplitude become constrained, 
$\sigma(c)\approx0.11$ with a 0.01 prior on $\om$ or 
$\sigma(c)\approx0.07$ with $\om$ fixed.  

This is an important point: analyses that do not 
take into account the physics of the form Eq.~\eqref{eq:pade} 
for scalar-tensor theories, but rather constrain 
an offset in the low $k$ (and often scale independent) 
limit, such as $\gm-1=\dg(z=0)(1+a^3)$, 
are addressing a very different physical model for 
gravity. Such constraints benefit 
greatly from the persistence effect: that any offset in 
$\gm$ at any redshift influences $\fs$ at all later 
redshifts. In contrast, the scalar-tensor form has 
little offset in $\gm$ at low $k$ at any redshift and 
so it is more difficult to constrain these physical models. 

\begin{figure}	
\includegraphics[width=\columnwidth]{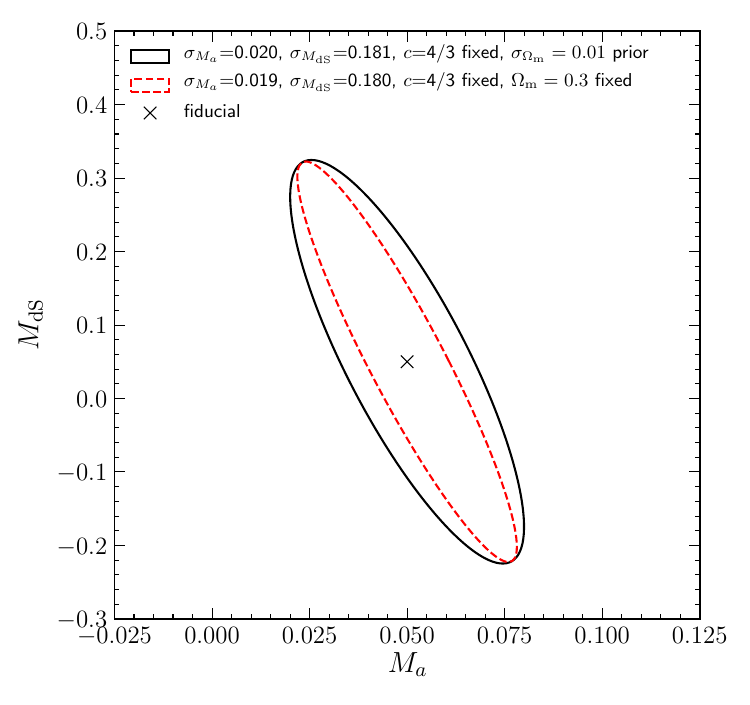}
\caption{
Joint 68\% confidence contours for $M_a$ and $\mds$ with fixed $c=4/3$, with a prior of 0.01 on $\om$ (solid curve) or fixed $\om$ (dashed curve). 
} 
\label{fig:pconstrf}
\end{figure}

To save the scalar-tensor form to some extent, we 
might fix $c=4/3$, the value for $f(R)$ theories. 
Here we give up some model independence and are now 
basically constraining $f(R)$ theories -- recall that 
the shape parameters $M_a$, $\mds$ are related to $B_0$ and 
$f_{R0}$ by roughly 
\be 
-f_{R0}\approx B_0/2 \approx \left[\frac{H_0}{M(a=1)}\right]^2 
\approx 10^{-5}\left(\frac{0.1}{M_a+\mds}\right)^2\ . 
\ee
We find constraints of $\sigma(M_a)=0.020$, $\sigma(\mds)=0.18$, 
insensitive to the $\om$ prior 
(fixing $\om$ gives $\sigma(M_a)=0.019$ and $\sigma(\mds=0.18)$). 
Recall that at high redshift, the $M_a$ 
term in Eq.~\eqref{eq:mofa} dominates 
over the $\mds$ term, and so is much better 
constrained. 
The joint 68\% confidence 
contour in the $M_a$--$\mds$ plane is shown in 
Fig.~\ref{fig:pconstrf}. 
Note that the larger $M_a$ and $\mds$ are, the 
closer the observables are to the GR case. 

One could propagate these uncertainties to the 
quantity $M_a+\mds$ to obtain an estimation 
uncertainty on $f_{R0}$. However, the Gaussian approximation of the information 
matrix approach is not robust here, giving $\sigma(M_a+\mds)/(M_a+\mds)\gtrsim1$. 
If one naively took that the $1\sigma$ 
upper limit on their sum is 0.27, this 
would imply a lower limit $|f_{R0}|\gtrsim 1.4\times 10^{-6}$. That is, for the given 
fiducial the data could distinguish the model 
from GR at 68\% CL. 
Note that the picture would be very different 
if one neglected the physics of the approach to 
the de Sitter state and ignored $\mds$. If the 
mass parameter were taken to be $M(a)=M_a a^{-3}$ only, 
then $\sigma(M_a)=\sigma(M(a=1))=0.01$ and one would 
conclude that $M_a=0.1\pm0.01$ (keeping the fiducial 
as $M(a=1)=0.1$) and therefore 
$|f_{R0}|=[0.8,1.2]\times 10^{-5}$ -- a very 
different physics conclusion! This highlights the 
importance of including all the key physics in 
interpreting the constraints from data.

\subsection{Binned Gravity Constraints} \label{sec:bin} 

To return to a more model independent approach to 
constraining gravity, we now estimate uncertainties of 
the $2\times3$ bin values of $\gm$, i.e.\ looking 
for trends between low and high $k$ (larger and 
smaller scales) and from high to mid to low redshift. 
The parameter set is now $(G_{11},G_{12},G_{13},G_{21},G_{22},G_{23},\om)$, 
and their fiducial values are the optimized fit 
values 
as described in Sec.~\ref{sec:binopt}. 

Table~\ref{tab:dgbin} compiles the results. At low $k$, 
the effect of the modified gravity on the observable $\fs$ 
is smaller than the precision of the data, and it is 
difficult to see any distinction from GR (i.e.\ $\gm=1$). 
For the $\keff=0.125$ bin, however, deviations from 
GR do start to have a noticeable effect. Note that 
since deviations in the low $a$ (high redshift) bin 
impact $\fs$ at all later redshifts, the precision on 
$\gm$ at low $a$ is tightest. At high $a$ (low redshift), 
the $\gm$ bin extending from $z=[0,0.5]$ only affects 
three data slices having reasonable precision: those 
at $z=0.25$, 0.35, 0.45, and so the uncertainties on 
$\gm$ at high $a$ become larger.

\begin{table} 
\centering 
\begin{tabular}{l|c|c|c} 
\hline 
Bin of $\gm$ & $G_{\rm bin}-1$  & $\sigma(G_{\rm bin})$ & $\sigma_+(G_{\rm bin})$\\ 
\hline 
\rule{0pt}{1.05\normalbaselineskip}low $k$, low $a$ [11]\quad{} & {}\quad\, 0.006 \quad\, & {}\quad\, 0.019 {}\quad\,&{}\quad\, 0.018 \quad\, \\ 
\rule{0pt}{1.05\normalbaselineskip}low $k$, mid $a$ [12]\quad & 0.031 & 0.104 & 0.103\\ 
\rule{0pt}{1.05\normalbaselineskip}low $k$, high $a$ [13] & 0.052 & 0.377 & 0.137\\ 
\rule{0pt}{1.05\normalbaselineskip}high $k$, low $a$ [21] & 0.029 & 0.018 & 0.018\\ 
\rule{0pt}{1.05\normalbaselineskip}high $k$, mid $a$ [22] & 0.127 & 0.094 & 0.093\\ 
\rule{0pt}{1.05\normalbaselineskip}high $k$, high $a$ [23]\quad{} & 0.158 & 0.290 &0.135\\ 
\hline 
\end{tabular} \\  
\caption{Information matrix constraints on $\gm$ in each 
of the $2\times3$ bins. Low/high 
$k$ refers to $\keff=0.055$, 0.125 respectively; low/mid/high 
$a$ refers to the ranges [0.37,0.5], [0.5,0.667], [0.667,0.833]. 
The $G_{\rm bin}-1$ column gives the fiducial value 
(i.e.\ the optimized values from Sec.~\ref{sec:binopt}) 
and the last two columns shows the $1\sigma$ uncertainties in 
determining those values from the projected redshift space distortion 
data ($\sigma$ column), and also including a peculiar velocity 
measurement ($\sigma_+$ column; see Sec.~\ref{sec:lowz}). 
} 
\label{tab:dgbin} 
\end{table}

\begin{figure*}	
\includegraphics[width=\textwidth]{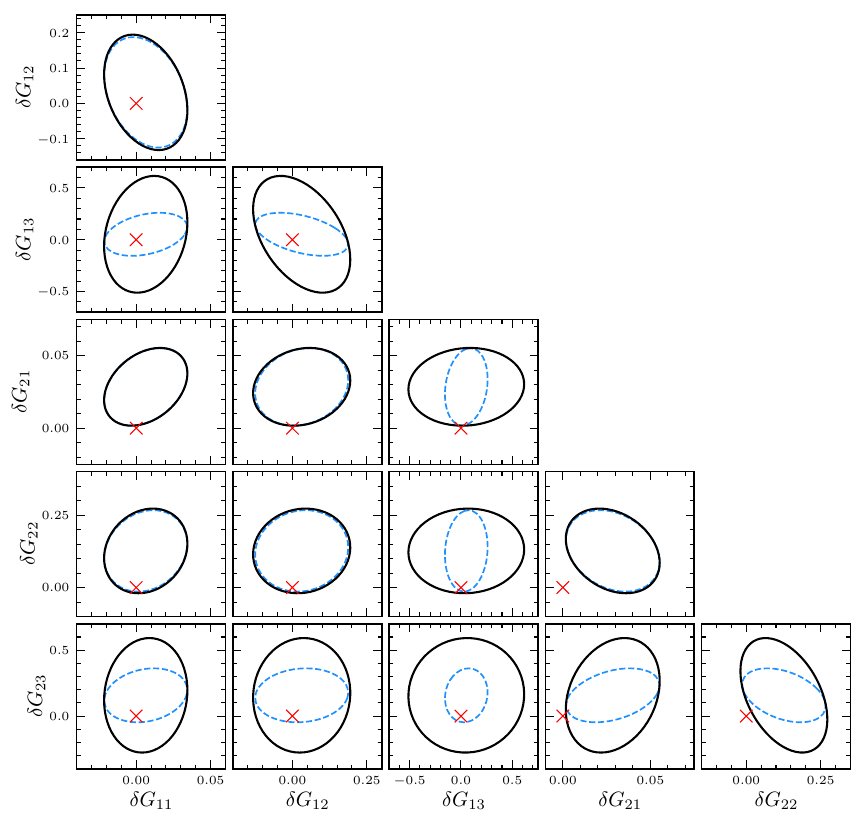}
\caption{Corner plot showing 2D joint 68\% CL 
constraints on the gravitational strength parameters 
in the two bins of $k$ and three bins of redshift. 
The red $\times$ denotes the GR value (0,0). 
The solid black contours use redshift space distortion 
data over $z=[0.2,1.7]$; the blue dashed contours add 
peculiar velocity data at $z=0.1$ (see Sec.~\ref{sec:lowz}). 
The matter density $\om$ is marginalized over with 
a prior of 0.01. 
	}
\label{fig:corner}
\end{figure*}

The power to discriminate modifications from GR 
becomes greater when assessing the joint, rather than 
1D, likelihood. Figure~\ref{fig:corner} shows the 
marginalized 2D joint confidence contours for all combinations 
of the $G_{\rm bin}$. Note that the parameters are 
fairly independent, with the greatest correlation 
coefficient being $-0.6$ between $G_{21}$ and $\om$; 
between the $G_{\rm bin}$ themselves, the maximum 
amplitude correlation is $-0.4$.

The red $\times$'s show the GR value in each panel, and 
we see that at low $k$ there is no statistically 
visible deviation from GR, even among the 
crosscorrelations (upper tip of the triangle). 
At high $k$ we see GR lies at or outside the 
68\% CL contour, and the deviation is especially 
noticeable in the crosscorrelations (right tip of the 
triangle). 
The overall difference in a GR fit to growth rate 
data from a cosmology with the fiducial modified gravity 
is $\Delta\chi^2=+20$, with 6 fewer parameters, giving 
at least a hint that one should look beyond GR. 
As in Sec.~\ref{sec:pade} we find that tightening the prior 
on $\om$ gives little improvement since 
there is not much covariance.  
Only a further generation 
of more precise data could improve the constraints on 
gravity in this model independent approach -- or 
combination with other probes of growth.

\subsection{Impact of Low Redshift Data} \label{sec:lowz} 

Another method being implemented with current surveys, 
such as DESI, to measure directly the growth rate is 
peculiar velocities \cite{1903.07652,1911.09121}. 
This can improve the precision at low redshift dramatically 
over redshift space distortions, which are limited by 
low available volume for clustering statistics. 
We consider the addition of a single 
redshift measurement 
for $\fs$ at $z=0.1$ with precision 3\% at each 
of $\keff=0.055$, 0.125 and investigate its impact on the 
gravitational constraints. This precision is 
a rough approximation of what DESI spectroscopy may 
provide from peculiar velocity measurements \cite{1903.07652,1911.09121}. 
(Note that in the future kinematic Sunyaev-Zel'dovich (kSZ) velocity measurements may also make precise growth rate constraints; see, e.g., \cite{1604.01382,2001.08608,2005.00523,2110.11127}.) 

This single redshift data point improves the constraints 
on the binned $\gm$ approach as shown by 
the last column in Table~\ref{tab:dgbin} for the 1D uncertainties and the dashed 
blue contours in Fig.~\ref{fig:corner} for the 2D joint constraints. Leverage is 
added particularly in the late time bin, i.e\ $G^{\rm bin}_{i3}$, 
such that GR is now pulled to the edge of many of the 2D 
joint confidence contours. The greatest effect is on 
$G_{13}$--$G_{23}$, where the uncertainty area is 
reduced by a factor 5.8. 
The total $\Delta\chi^2$ distinction 
from GR is now $\Delta\chi^2=28$, 
i.e.\ the one peculiar velocity data point 
distinguishes from GR by $\Delta\chi^2=8$. 

\begin{figure}	
\includegraphics[width=\columnwidth]{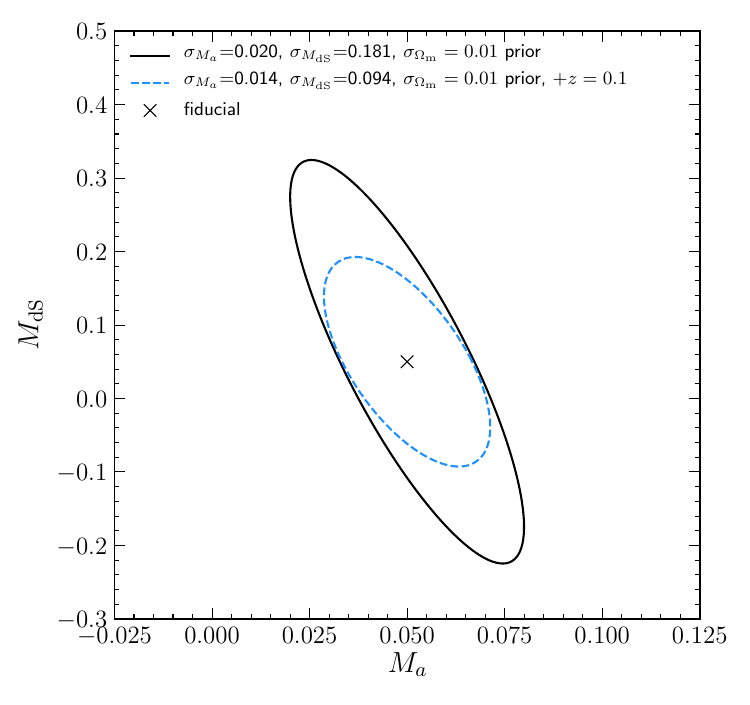}
\caption{
Joint 68\% confidence contours for $M_a$ and $\mds$ with fixed $c=4/3$ and a prior of 0.01 on $\om$, for redshift 
space distortion only (solid curve, as 
in Fig.~\ref{fig:pconstrf}), and with the peculiar velocity data at $z=0.1$ (dashed blue curve). 
} 
\label{fig:mmpv}
\end{figure}

For the scalar-tensor Pad{\'e} approach 
we also find that the peculiar velocity data 
on the growth rate at $z=0.1$ can have good 
complementarity with the redshift space 
distortion measurements. While this one added 
data point does help to determine the amplitude 
parameter $c$, lowering $\sigma(c)$ from 1.7 
without the $z=0.1$ data to 0.8 with, this is still 
not sufficient. Fixing $c=4/3$ as for $f(R)$ theories 
as discussed in Sec.~\ref{sec:pade}, the peculiar 
velocity data has excellent leverage, as seen 
in Fig.~\ref{fig:mmpv}. Now $\sigma(M_a)=0.014$, 
down from 0.020, and $\sigma(\mds)=0.094$, down from 
0.181. The overall uncertainty area reduces by a 
factor 2 by adding the $z=0.1$ peculiar 
velocity measurement. Using the same naive procedure 
as before to estimate $f_{R0}$, the $1\sigma$ 
upper limit on $M_a+\mds$ is 0.19, for a 
lower limit on the fiducial model of 
$|f_{R0}|\gtrsim 2.9\times 10^{-6}$.

\section{$2\times2$ Bins} \label{sec:bin22} 

To get clearer distinction from GR we could 
use fewer bins in $a$, albeit at the loss 
of some evolutionary information on $\gm$. 
To explore this, we consider two bins in $a$ 
for $\gm$: $a=[0.25,0.556]$ and $a=[0.556,0.833]$, 
with the dividing line at $z=0.8$ so there will 
be sufficient precise data in both bins. 
The two bins in $k$ remain as is, with a 
piecewise constant form for the $\gm$ bin values. 
For just two bins in $a$, piecewise constant 
$\gm$ will not give a sufficiently accurate 
$\fsh$ relative to the exact $\fs$, so we 
use a piecewise linear form, with continuity 
at the bin boundary $z=0.8$. 
(See Sec.~\ref{sec:binopt} for a discussion of bins vs interpolated node values. In actual data analysis one might want to assess both.) 
We call our 
new $2\times2$ gravity parameter set $G_{1A}$, 
$G_{1B}$, $G_{2A}$, $G_{2B}$, where the first 
index indicates low/high $k$ as before, and 
the second index $A/B$ indicates low/high $a$. 
(We change from numbers to letters to avoid 
confusion with the $G_{11}$ etc.\ from the 
$2\times3$ set.) The bin values are defined 
at the midpoint in $a$ of each bin. 

\begin{figure}	
\includegraphics[width=\columnwidth]{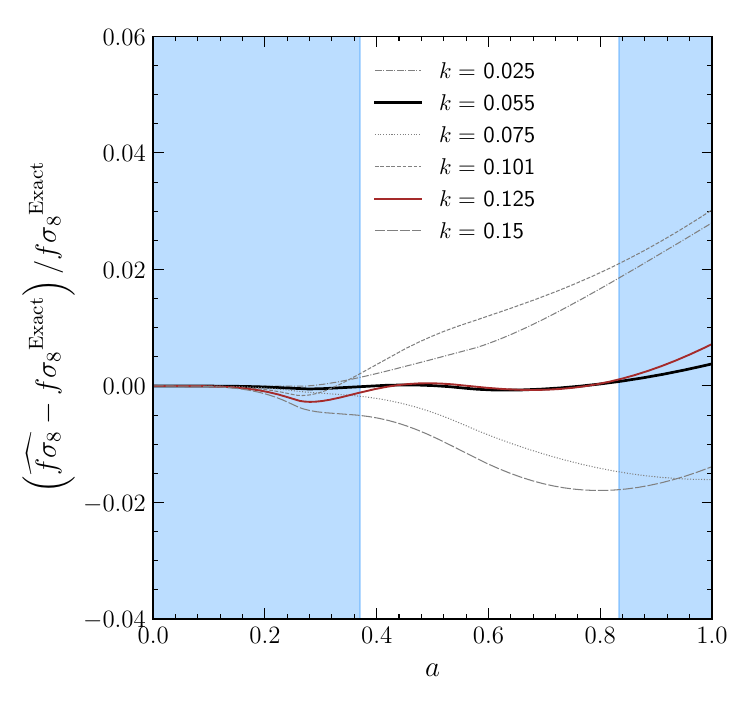} 
\caption{The fractional difference of $\fs$ using the 
optimized $2\times 2$ binned $\gm^{\rm bin}$ (giving $\fsh$) vs 
using the exact modified gravity $\gm$ 
(giving $\fs^{\rm Exact}$) is plotted vs $a$ for various 
$k$. 
As in Fig.~\ref{fig:dfsgrbe}, only the values at $\keff=0.055$, 
0.125 are observationally relevant. 
Similarly, the relevant (i.e.\ good 
data precision) 
range in $a$ is indicated by the unshaded region. 
	}
\label{fig:dfsgrbe2}
\end{figure}

Figure~\ref{fig:dfsgrbe2} illustrates that 
the maximum deviation in reconstructing $\fs$ 
can be under 0.2\% even with only two bins in 
$a$. As before we use bin centers at $\keff=0.055$, 
0.125 and optimize the values of $\dg^{\rm bin}$ 
to minimize the rms fractional deviation of 
$\fsh$ from $\fs$ given by the fiducial scalar-tensor 
model. The optimized values are $\dg_{1A}=0.011$, 
$\dg_{1B}=0.048$, $\dg_{2A}=0.049$, $\dg_{2B}=0.15$.

\begin{figure*}	
\includegraphics[width=\textwidth]{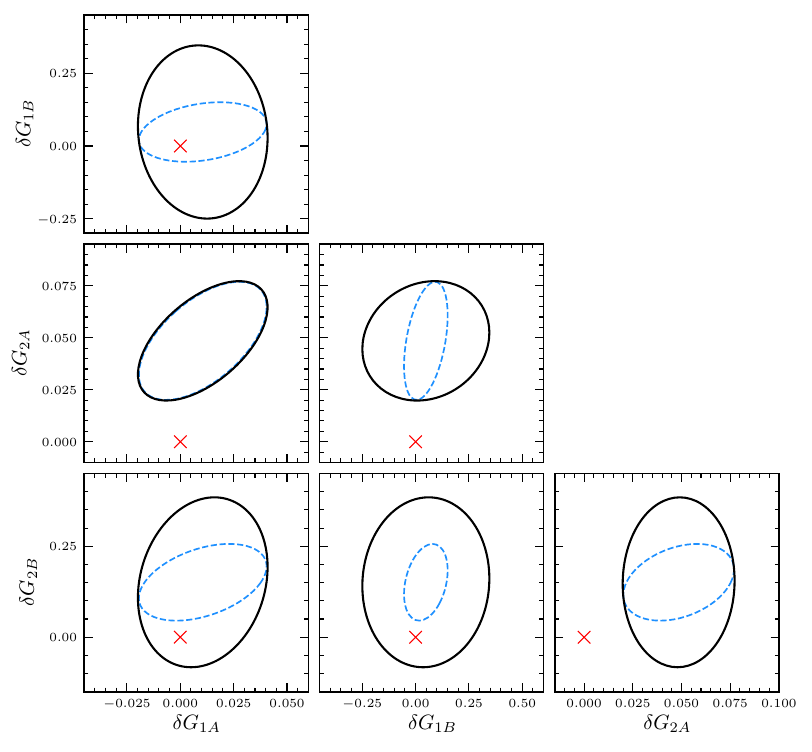}
\caption{Corner plot showing 2D joint 68\% CL 
constraints on the gravitational strength parameters 
in the two bins of $k$ and two bins of redshift. 
The red $\times$ denotes the GR value (0,0). 
The solid black contours use redshift space distortion data over $z=[0.2,1.7]$; the dashed blue contours add peculiar velocity data at $z=0.1$ (see Sec.~\ref{sec:lowz}). 
The matter density $\om$ is marginalized over with a prior of 0.01. 
	}
\label{fig:corner22}
\end{figure*}

Therefore we can use the $2\times2$ bin 
parametrization for robust comparison to 
projected data, and derive constraints on the 
gravity parameters. The corner plot of the 
2D joint confidence contours is shown in 
Fig.~\ref{fig:corner22}. 
Now, several panels show GR as lying 
significantly outside the 68\% confidence 
contour, either using redshift space 
distortion data alone or adding the 
peculiar velocity data at $z=0.1$. In some 
cases it lies outside the 95\% joint 
confidence limit. 

As before, 
the discrimination from GR is most 
statistically significant for the higher 
$k$ bin. 
The higher redshift bin also has tighter 
constraints than the lower redshift bin. 
The peculiar velocity data 
continues to exhibit strong leverage in 
complementarity with redshift space 
distortion measurements. In the $2\times2$ 
case, the total distinction from GR is 
now $\Delta \chi^2=23$, for two fewer 
parameters than the $2\times3$ case. 
This improves to 
$\Delta \chi^2=32$ with the peculiar velocity 
data point at $z=0.1$. The greatest 
improvement due to adding peculiar 
velocity data is on the $\dg_{1B}-\dg_{2B}$ 
joint contour, where the overall uncertainty 
area is reduced by a factor of 7.

\section{Conclusions} \label{sec:concl} 

The growth rate of cosmic structure is now 
being measured precisely by the current generation 
of spectroscopic surveys. This can combine with 
expansion history information to tighten constraints 
on dark energy and cosmic acceleration. However, 
beyond general relativity the growth rate carries 
extra information on the theory of gravitation, and 
for many theories this includes scale dependence 
even in the linear to quasilinear density regime. 

We have demonstrated that the scale dependence in 
canonical scalar-tensor theories of gravity can be 
accurately accounted for by a model independent 
binning of the gravitational coupling: 2 bins in 
$k$ and 3 bins in $a$ deliver 0.02-0.27\% rms 
accuracy in the full growth rate $\fs(k,a)$ even for  
a fairly conservative scalar-tensor gravity theory for 
comparison to observations. Such model independent 
bins have little covariance and are readily 
interpretable. 

If one does not use all the physics present in  
scalar-tensor gravity, e.g.\ uses a power law 
in $a$ for $\gm$ rather than the true Pad\'e form, 
or neglects the freezing of the scalaron mass 
as the universe is dominated by cosmic acceleration, 
we show that the results will be strongly different, 
both quantitatively and in interpretation. 
Some other models of gravity may contain a constant  
offset, i.e.\ constant deviation from GR even at 
low $k$, and these tend to be easier to constrain 
but address different physics. 

Fitting parameters in the scalar-tensor form itself, 
thus being somewhat more model dependent, 
does not work due to strong covariances. Only 
when a particular class is chosen, such as $f(R)$ 
gravity, to fix the amplitude parameter $c$ can the 
shape parameters $M_a$, $\mds$ describing the scale 
dependence be fit with reasonable precision. These 
can be related to more specific quantities such as 
$|f_{R0}|$ and we demonstrate potential distinction 
from general relativity by obtaining a lower bound 
$|f_{R0}|\gtrsim1.4\times 10^{-6}$ (in contrast to 
a spuriously tight $[0.8,1.2]\times 10^{-5}$ when 
ignoring the necessary de Sitter physics) for a 
model with true $|f_{R0}|\approx 10^{-5}$. 

Of great promise is the strong complementarity 
of low redshift peculiar velocity probes of the 
growth rate with redshift space distortions from 
surveys at higher redshift. Recall that the growth 
rate at low redshift is sensitive to the cosmic 
gravitational history at all higher redshifts. 
Peculiar velocity data at $z\approx0.1$ can 
tighten the 2D joint uncertainty area on binned 
gravity parameters by up to a factor 7, and 
raise the statistical significance of discrimination 
from general relativity to several measurements 
of 95\% confidence level, or an overall 
$\Delta\chi^2=28$. 
While the precisions we adopted for RSD and peculiar velocity measurements are reasonable projections, it is quite exciting that DESI has already acquired substantial data and that actual analysis, including for various $\kmax$ and with neutrino mass, will be forthcoming in the next few years. 

Finally, we also explore an 
alternate model independent parametrization 
with $2\times2$ bins of $\gm$ and find it 
also works well. 
It accurately reproduces the full scalar-tensor 
prediction for $\fs(\keff,a)$ with two fewer 
parameters. The constraints from projected data 
tighten, with greater discrimination power from 
GR. Two downsides are a more coarse grained view 
of the evolution in $a$, and a somewhat greater 
correlation between bins (because of the needed 
continuous piecewise linear rather than piecewise 
constant parametrization). One could envision 
using $2\times2$ as an alert for a deviation from 
GR, and then reanalyzing the data with $2\times3$ 
to pick up finer detail.

\acknowledgments 

This work was supported in part 
by the Energetic Cosmos Laboratory. 
EL is supported in part by the U.S.\ Department of Energy, Office of Science, Office of High Energy Physics, under contract no.\ DE-AC02-05CH11231. 

\bibliography{scalegrow}

\end{document}